\documentclass[conference]{IEEEtran}

\usepackage[T1]{fontenc}

\usepackage{amssymb}	 
\usepackage{amsmath}

\usepackage{graphicx}
\usepackage{booktabs}
\usepackage{multirow}
\usepackage{tabularx}
\usepackage{threeparttable}
\usepackage{url}

\newcommand{\vcs}{\textup{\textbf{\textrm{s}}}}
\newcommand{\mts}{\textup{\textbf{\textrm{S}}}}
\newcommand{\vcv}{\textup{\textbf{\textrm{v}}}}
\newcommand{\vcw}{\textup{\textbf{\textrm{w}}}}

\begin{document}

\title{An Empirical Analysis of  Vulnerabilities \\ in  Python Packages for Web Applications}
\author{
\IEEEauthorblockN{Jukka Ruohonen}
\IEEEauthorblockA{University of Turku, Finland \\ 
Email: juanruo@utu.fi}
}

\maketitle

\begin{abstract}
This paper examines software vulnerabilities in common Python packages used particularly for web development. The empirical dataset is based on the PyPI package repository and the so-called Safety DB used to track vulnerabilities in selected packages within the repository. The methodological approach builds on a release-based time series analysis of the conditional probabilities for the releases of the packages to be vulnerable. According to the results, many of the Python vulnerabilities observed seem to be only modestly severe; input validation and cross-site scripting have been the most typical vulnerabilities. In terms of the time series analysis based on the release histories,  only the recent past is observed to be relevant for statistical predictions; the classical Markov property holds.
\end{abstract}


\section{Introduction}

The patching of vulnerabilities is a classical topic. To provide means for prioritizing vulnerabilities to be patched, one common approach has built upon the questions of whether exploits for the vulnerabilities are known to exist (somewhere), and whether exploitation is observed to occur in the wild~\text{\cite{Almukaynizi17, Nappa15}}. This approach may be useful for operating system vendors, maintainers of large networks and cloud platforms, and to  national organizations trying to inform consumers about the necessity of timely patching. Yet, the questions related to exploits and exploitation are quite far from the current software engineering settings with which typical software is continuously developed, deployed, and maintained.

Development, testing, integration, and other software engineering activities currently often occur on different cloud computing platforms. In addition to continuous testing, integration, and delivery, automation is one of the contemporary keywords~\cite{Visser18}. The automated continuous deployment frameworks are important for vulnerability monitoring particularly in the web application domain. Most current web applications are not only developed but also deployed on cloud platforms by using containers, virtual machines, and related solutions. Systematic monitoring of vulnerabilities within such deployment systems is a good example of the new practical challenges~\cite{Plate15, Schleicher16}. These transformations in the ways software is engineered have also complicated the tracking of known vulnerabilities. 

Good progress has recently been made to better understand the disclosure and patching of software vulnerabilities in open source projects \cite{LiPaxson17, Ruohonen18IST}. However, much of the research is based on the Common Vulnerabilities and Exposures (CVEs). While there are good and justifiable reasons to focus on CVE-stamped vulnerabilities, this focus also limits the scope of a viewpoint; not all known vulnerabilities have CVEs. In fact, the absence of CVE identifiers for many vulnerabilities has long been the message from those promoting alternative vulnerability databases~\cite{Dunn18, Pyup18}. In the open source context these identifiers are generally allocated only for well-coordinated and well-understood vulnerabilities in important open source software packages. Less important, poorly maintained, or even  obscure open source packages are often bypassed---and hosting services such as GitHub are full of such packages.

These limitations have prompted a new branch of research for examining vulnerabilities in software repositories. While packages used in Linux distributions have been a common target~\cite{Stuckman14}, the more recent research has focused on language-specific repositories such as \textit{npm} for JavaScript \cite{Decan18a, Ruohonen18TIR}. This is the research domain to which this paper contributes by presenting the supposedly first study on vulnerabilities in the Python's PyPI repository and advancing the understanding on release-based time series analysis of software vulnerabilities.

\section{Data}\label{section: data}

\subsection{Sources}\label{subsec: sources}

The dataset used is based on three sources. The primary data source is the so-called Safety~DB that was launched in 2016 for tracking vulnerabilities in open source packages written in Python~\cite{SafetyDB18}. The packages tracked in the database are mostly related to web application development. The secondary data source is the so-called Python Package Index (PyPI) used to archive and manage Python projects~\cite{PyPI18}. In addition, a few descriptive observations rely on further data fetched from NVD, that is, the National Vulnerability Database \cite{NVD18b}. The following cut JSON (JavaScript Object Notation) excerpt can be used to illustrate a rather typical entry in Safety DB:

\begin{small}
\begin{verbatim}
{ "advisory": "The django.views.static.serve 
   view in Django before 1.4.18, 1.6.x before 
   1.6.10, and 1.7.x before 1.7.3 reads [...]"

  "cve": "CVE-2015-0221", "id": "pyup.io-33072",
  
  "specs": [ "<1.4.18",
             ">=1.6,<1.6.10",
             ">=1.7,<1.7.3" ],
  
  "v": "<1.4.18,>=1.6,<1.6.10,>=1.7,<1.7.3" }
\end{verbatim}
\end{small}

As can be seen, the \texttt{advisory} field provides a brief textual description for each vulnerability archived to the database. These descriptions follow the typically terse prose used for describing vulnerabilities \cite{Ruohonen18TIR}. (It is also worth remarking that the textual advisories in Safety DB are mostly plagiarized directly from NVD and related sources.) To accompany the textual descriptions, the \texttt{cve} field records a CVE potentially assigned for a vulnerability archived to the database. The more valuable information is stored to the \texttt{specs} and \texttt{v} fields, which both define the releases affected by a vulnerability. 

\subsection{Operationalization}\label{subsec: operationalization}

The release-based dataset was assembled as follows. For each package in Safety DB, the full available release history was first retrieved from the PyPI repository with the \textit{pip} package manager. Then, for each vulnerability, a zero-valued vector ${\vcv = [ 0, \ldots, 0 ]}$ was initialized such that the length $r$ of the vector equals the length of the corresponding package's ordered release history. The first element in a given $\vcv$ thus corresponds with the first release made for the given package.

The vectors were filled according to the release specifications provided in Safety DB. To process the specifications, a given version in the \texttt{specs} array was parsed only if the version was present in a package's release history reported in PyPI. For the interval-based specifications---such as the \texttt{">=1.6,<1.6.10"} condition used in the previous excerpt, both versions were required to have valid entries. Unlike in some previous studies \cite{Decan18b}, pre-releases were not filtered out. Therefore, the version sequences may contain also alpha releases (cf.~\texttt{1.2.3-alpha}), beta releases, release candidates (cf.~\texttt{1.2.3-rc.0}), and other abstract release types that do not fit directly into the so-called semantic versioning scheme. In any case, for each specification, a zero-valued $r$-length vector $\vcs$ was again initialized, and then filled with the value one for each release affected. The binary AND operator was used for filling the interval specifications. As an example: if
\begin{equation}\label{eq: left-right example}
\begin{cases}
\vcs_\textmd{left} &= [ 0, 0, \ldots, 0, 1, 1, 1, 1, \ldots, 1] \\
\vcs_\textmd{right} &= [ 1, 1, \ldots, 1, 1, 1, 0, 0, \ldots, 0]
\end{cases}
\end{equation}
denote two filled vectors for an interval specification, the resulting vector $\vcs$ would be $[ 0, 0, \ldots, 0, 1, 1, 0, 0, \ldots, 0]$. The per-vulnerability specification vectors were then collapsed into a given $h \times r$ matrix $\mts$, where $h$ denotes the number of specification vectors for a given vulnerability.  The binary OR operator was then used to construct a given vector $\vcv$. That is:
\begin{align}
\vcv = 
\left[
\bigvee^h_{i=1} s_{i1}, \ldots,
\bigvee^h_{i=1} s_{ir} 
\right]
~\textmd{given}~s~\textmd{in}~\mts.
\end{align}

The final operationalization step involved per-package aggregation. If a given package was affected by $m$ vulnerabilities in total, such that $\vcv_1, \ldots, \vcv_m$ vectors were filled for the package, simple summation was first used to calculate the number of times a given release of the package was affected:
\begin{equation}\label{eq: count}
\tilde{\vcw} = 
\left[ 
\sum^m_{i=1} v_{1_i} ,
\ldots ,
\sum^m_{i=1} v_{r_i} 
\right]
~\textmd{given}~v~\textmd{in}~\vcv_k 
\end{equation}
and $~k = 1, \ldots, m $. These count data vectors are not ideal for empirical analysis, however. The reason relates to the commonplace industry saying that ``stop counting vulnerabilities''. 

The essence behind the saying is that vulnerability counts do not necessarily matter in terms of actual security. The point extends to maintenance: practical package management seldom concerns the question of how many times a given release has been affected by vulnerabilities---a single vulnerability is enough to upgrade. Furthermore, vulnerability counts cannot be reliably used to compare products, packages, or other software elements. This point is also emphasized by the maintainers of Safety~DB---the database ``is not a hall of shame, or a list of packages to avoid''~\cite{SafetyDB18}. If the database would be used as a blacklist, the resulting list would contain well-maintained web application projects such as Django, whereas some smaller projects might be shortlisted as safe because these are often less rigorously maintained. For these and related reasons, binary-valued vectors are actually observed for all packages: an element $w$ in an observed vector $\vcw$ takes a value zero for a fully vulnerability-free release and a value one if $\tilde{w} > 0$ within a corresponding vector $\tilde{\vcw}$ in~\eqref{eq: count}.

\section{Results}

\subsection{Overview}

A few general observations are presented before the time series results. The subset of CVE-referenced vulnerabilities (about 31\% of all vulnerabilities) helps to deliver these points. 

\subsubsection{Calendar time}

\begin{figure}[th!b]
\centering
\includegraphics[width=\linewidth, height=2.9cm]{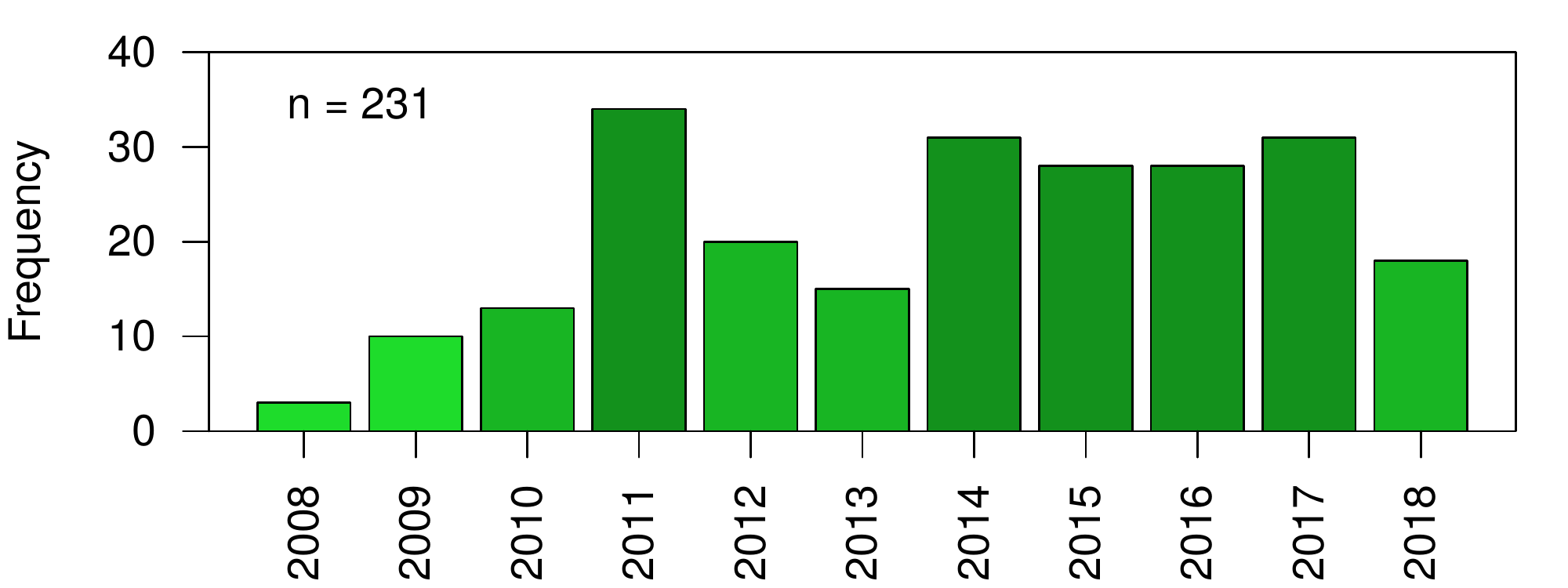}
\caption{Publication Years (NVD) of the CVE-referenced Vulnerabilities}
\label{fig: years}
\end{figure}

It is necessary to point out that Safety~DB does not record any calendar time information for the vulnerabilities archived. By using the dates on which the CVE-referenced vulnerabilities were published in NVD, it seems that the vulnerabilities were mostly discovered and disclosed during the 2010s (see Fig.~\ref{fig: years}). By further examining the database's commit history, it becomes evident that the database was in fact initially built by examining NVD and related sources. The later entries are based on semi-manual keyword-based monitoring of commit logs and bug trackers in GitHub~\cite{Pyup18}. In contrast to the so-called Snyk database~\cite{Ruohonen18TIR}, no references are provided for the commits or bug reports, however. This lack of references implies that it is impossible to reconstruct the calendar time events. Therefore, it should be also noted that an analysis cannot rely on  analytical operators and theoretical concepts sometimes used to examine release histories~\cite{Decan18b}. But as will be elaborated, the binary-valued vectors still provide a decent source for many insights even when the calendar time intervals between releases are ignored.

\subsubsection{Squatting}

The so-called typo-squatting has become a bugbear for open source software package repositories. Particularly \textit{npm} has been affected by these simple squatting attacks with which attackers upload malicious packages to a repository with highly similar names as the names of already existing legitimate packages. Also PyPI has been targeted, first in September 2017~\cite{Claburn17} and then again in October 2018~\cite{ZDNet18}. Although these cases are excluded from the analysis, it is worth noting that both typo-squatting campaigns were likely serious attempts; among the names of the malicious packages were \texttt{telnet}, \texttt{djanga}, \texttt{crypt},
\texttt{urlib3}, and \texttt{setup-tools}. 

\subsubsection{Severity}

\begin{figure}[th!b]
\centering
\includegraphics[width=\linewidth, height=2.9cm]{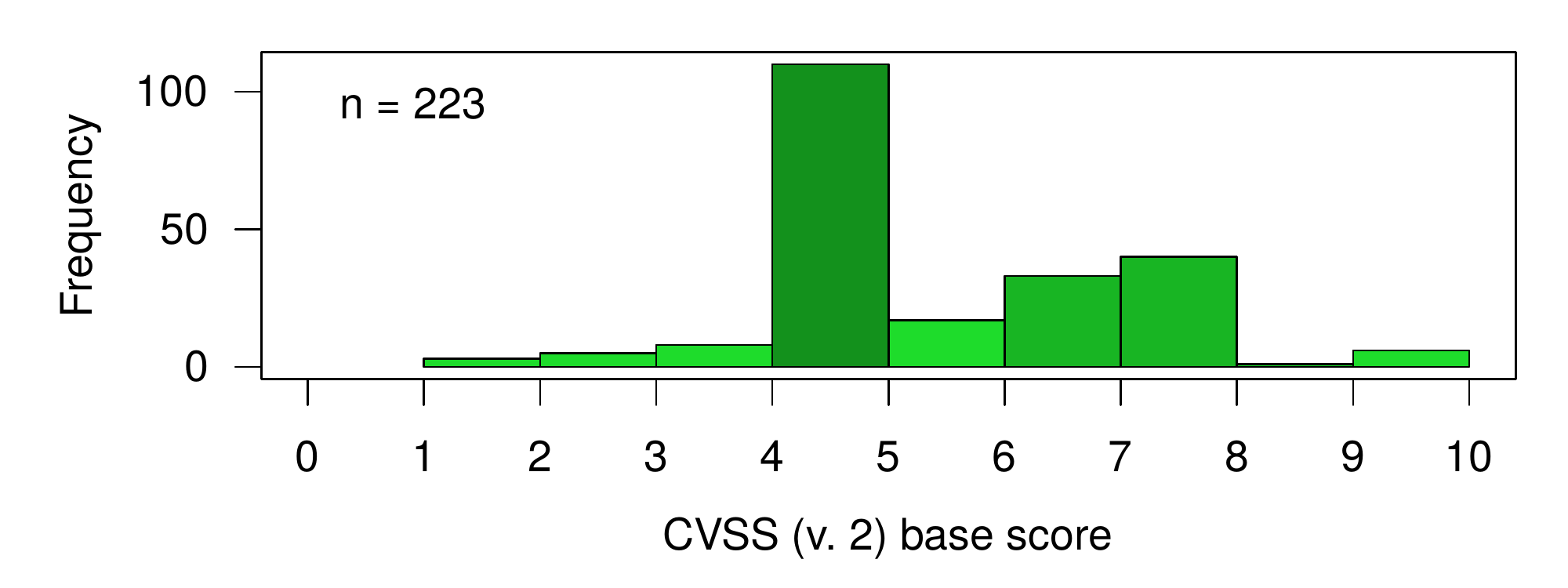}
\caption{The Severity of the CVE-referenced Vulnerabilities}
\label{fig: cvss}
\end{figure}

The Common Vulnerability Scoring System (CVSS) is widely used to quantify the severity of vulnerabilities. For obtaining a quick overview, the so-called CVSS~(v.~2) base scores are sufficient. These range from zero to ten; the higher the score for a vulnerability, the more severe the vulnerability. To aid the interpretation, it can be mentioned that the value four is a common but somewhat arbitrary benchmark for the base scores with respect to the importance of patching~\cite{Allodi17b}. Given these remarks, the base CVSS scores are shown in Fig.~\ref{fig: cvss} for the intersecting subset of vulnerabilities in Safety~DB that have CVEs and the corresponding vulnerabilities in NVD that have CVSS entries. Presumably due to the small delays in NVD's CVSS vulnerability scoring~\cite{Ruohonen18ACI}, it can be remarked that eight recent CVE-referenced vulnerabilities in Safety~DB did not yet have CVSS data in NVD at the time of writing.

The subset of vulnerabilities with both CVE and CVSS information indicate only relatively modest severity. The median base score is five. While the histogram shown partially reflects the second version of the CVSS standard and the general distribution in NVD~\cite{Gallon11b}, it also reflects the nature of Python as an interpreted language and the typical software packages written with the language. Severe low-level vulnerabilities are less common in this context. This said, there are two vulnerabilities with the maximum score: CVE-2016-4009 and CVE-2016-5636. Both refer to heap-based buffer overflows that can be exploited remotely. The corresponding packages are Pillow (a common library for handling images in Python) and CPython, that is, the reference yet the \textit{de facto} interpreter for the whole language. These two cases are sufficient to underline that common packages for interpreted languages are not invulnerable to memory corruption issues. Many packages are written with multiple programming languages, and programming languages are written with programming languages.

\subsubsection{Weaknesses}

Software weaknesses are another abstraction related to vulnerabilities. When used with already known vulnerabilities, these abstractions convey the underlying programming mistakes and software development flaws behind the concrete vulnerabilities observed. The so-called Common Weakness Enumeration (CWE) framework is used in NVD and related databases for cataloging the weaknesses behind the vulnerabilities archived to the databases. Although Safety~DB does not use CWE, a subset of CVE-referenced vulnerabilities that have CWE entries in NVD provides a decent glimpse on the typical software weaknesses affecting Python packages.

\begin{figure}[th!b]
\centering
\includegraphics[width=\linewidth, height=7.9cm]{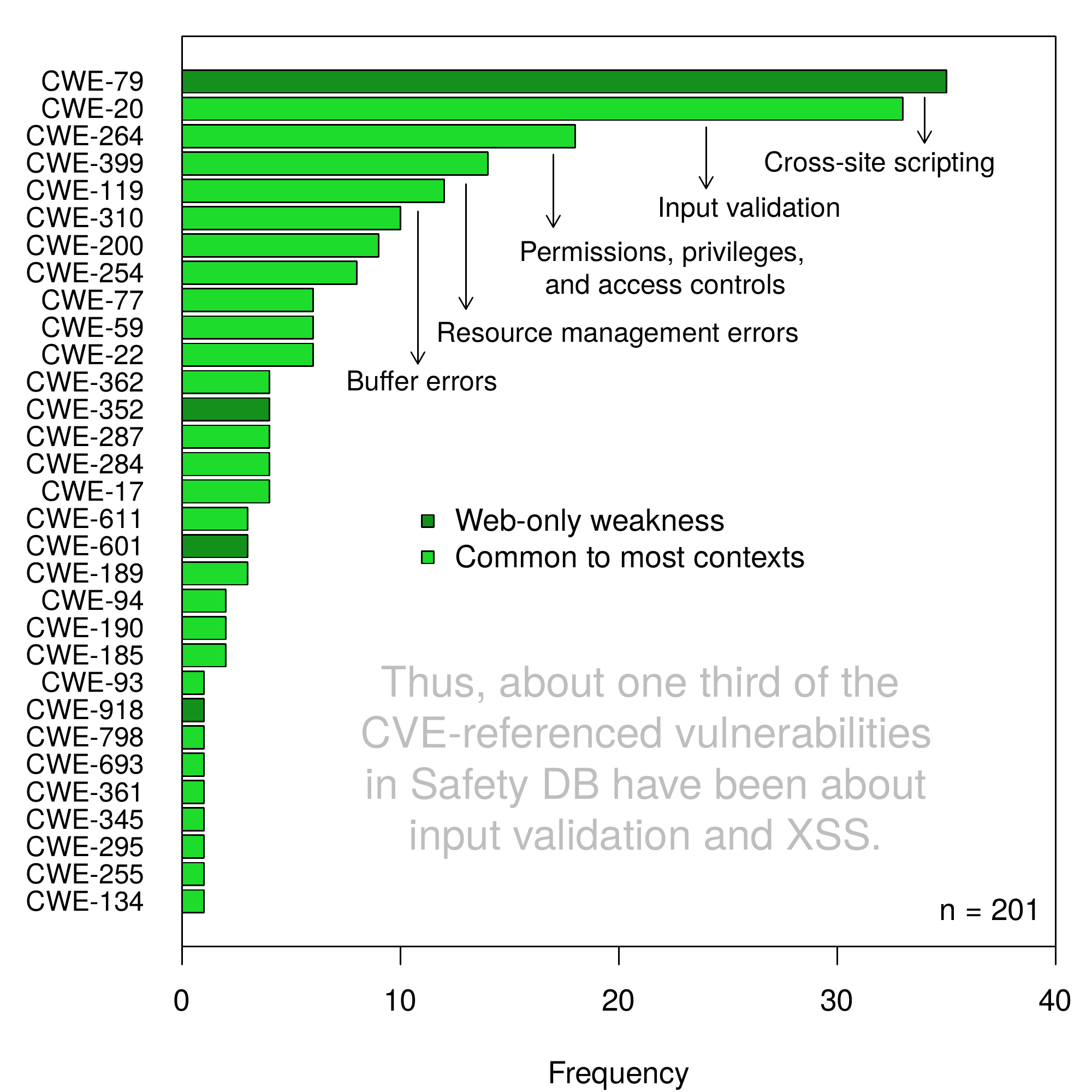}
\caption{The Weaknesses Behind the CVE-referenced Vulnerabilities}
\label{fig: cwe}
\end{figure}

The results summarized in Fig.~\ref{fig: cwe} show no big surprises in terms of the overall distribution of contemporary software vulnerabilities; cross-site scripting (XSS) vulnerabilities and more generally input validation bugs have continued to occupy many top-rankings. One explanation for the explosion of cross-site scripting vulnerabilities in the 2000s related to the adoption boom of the PHP programming language and, perhaps to a lesser extent, JavaScript's increased popularity~\cite{Sivakumar07}. According to the results, also the later increase in Python's popularity~\cite{Cass18} seems to have contributed to the prevalence of XSS bugs. Even though also resource management flaws and buffer-related weaknesses have been unexpectedly common in the web-related Python packages observed, the only real surprise is the complete absence of structured query language (SQL) injection bugs (CWE-89) in the CVE-referenced subset.

\subsection{Releases}

There were $765$ vulnerabilities in Safety DB at the time of retrieving the database. Due to the constraints discussed in Subsection~\ref{subsec: operationalization}, this amount reduced to $526$ vulnerabilities that affected the releases of $335$ packages. The following results are based on the binary vectors $\vcw_1, \ldots, \vcw_{335}$ constructed.

\subsubsection{Overview}

\begin{figure}[th!b]
\centering%
\includegraphics[width=\linewidth, height=4.0cm]{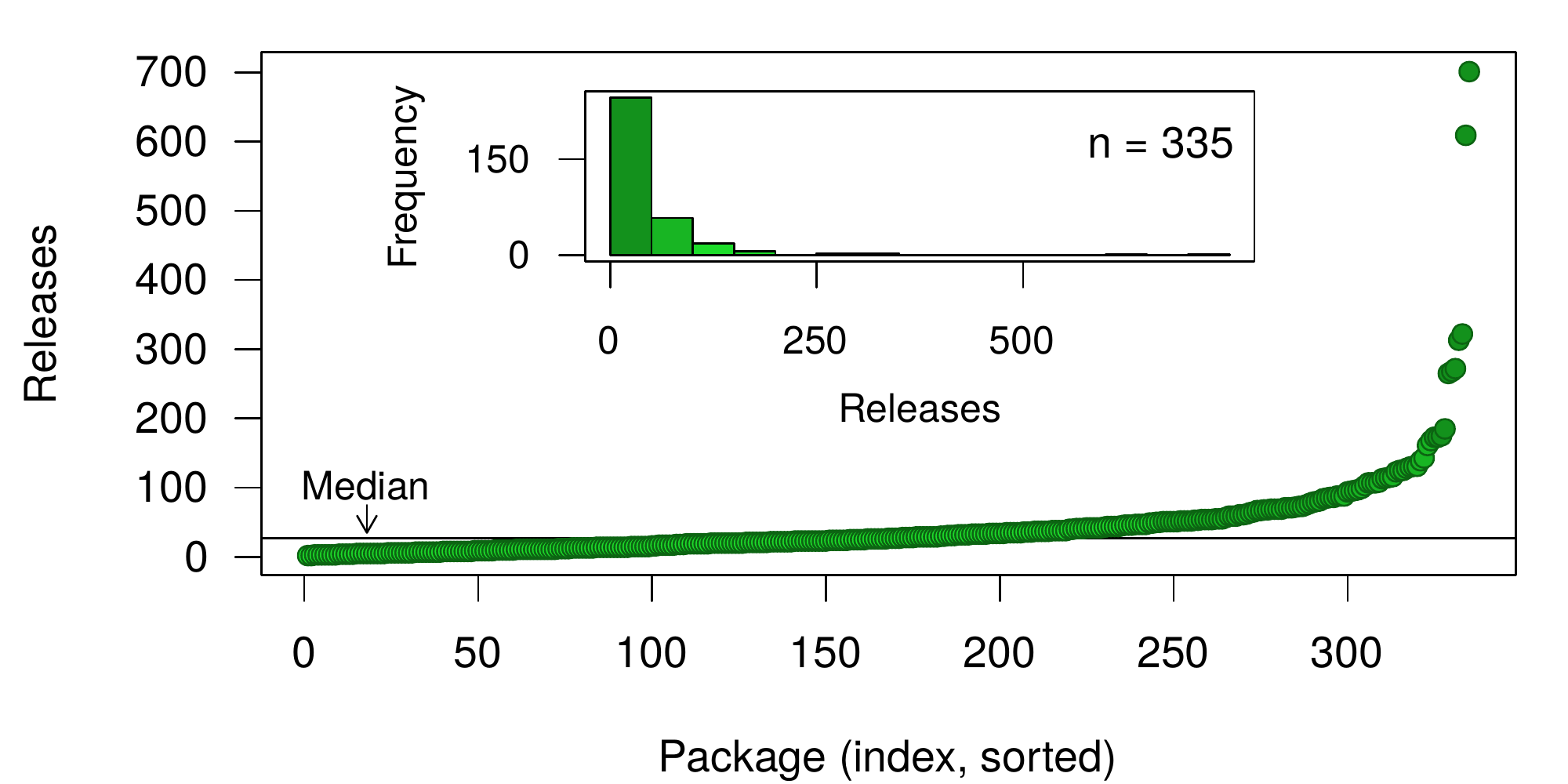}
\caption{Number of Releases According to PyPI}
\label{fig: releases}
\end{figure}

Before turning to the actual vulnerabilities, it is useful to take a brief look at the release histories of the packages observed. The lengths of the per-package release histories are visualized in Fig.~\ref{fig: releases}. Although the standard deviation is substantial, many of the packages have a relatively long release lineage; the median is $27$ releases and the 75th percentile is a little over fifty releases. This observation hints that the packages tracked in Safety~DB are generally mature and supposedly relatively well-maintained. While repository-wide release engineering strategies and associated monitoring may be desirable~\cite{Decan18b}, the two visible outliers (\texttt{awscli} and \texttt{pytsite}) are enough to point out that the current release engineering practices vary a lot between Python packages.

\begin{figure}[th!b]
\centering%
\includegraphics[width=\linewidth, height=3.9cm]{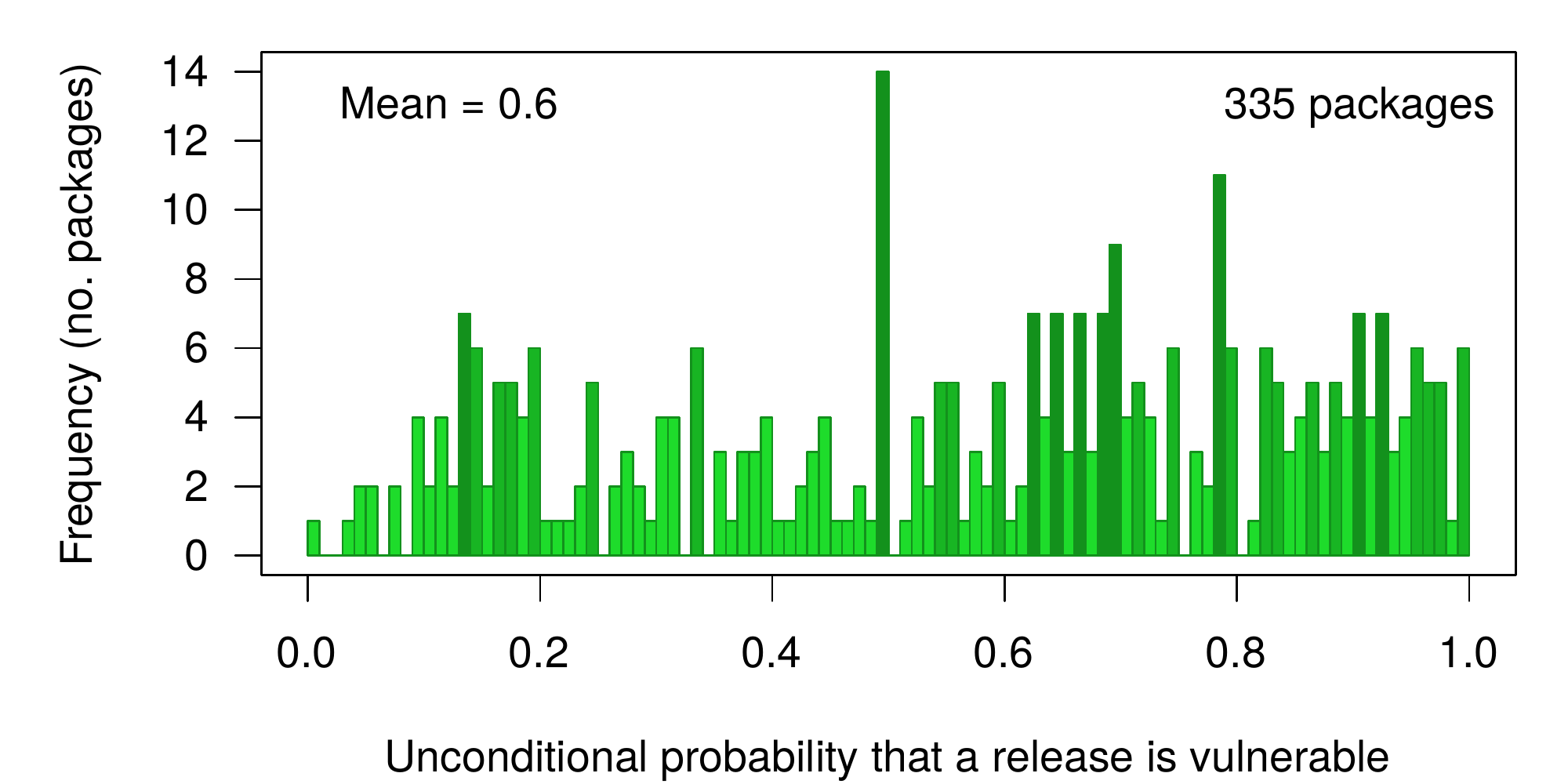}
\caption{Unconditional Probabilities}
\label{fig: uncond}
\end{figure}

A basic question for release-based vulnerability analysis is how common it is for a release to be vulnerable. If there is no other available information whatsoever, this question amounts to calculating the sum of a $\vcw_k$ divided by the vector's length. The results are shown in Fig.~\ref{fig: uncond} for all of the vectors. The unconditional probability is quite large for the packages' arbitrarily picked releases to be vulnerable. On average over a half of the releases observed have been affected by at least one vulnerability. The median is $0.6$. The shown distribution across the packages resembles the uniform probability distribution.

\begin{figure}[th!b]
\centering%
\includegraphics[width=\linewidth, height=5cm]{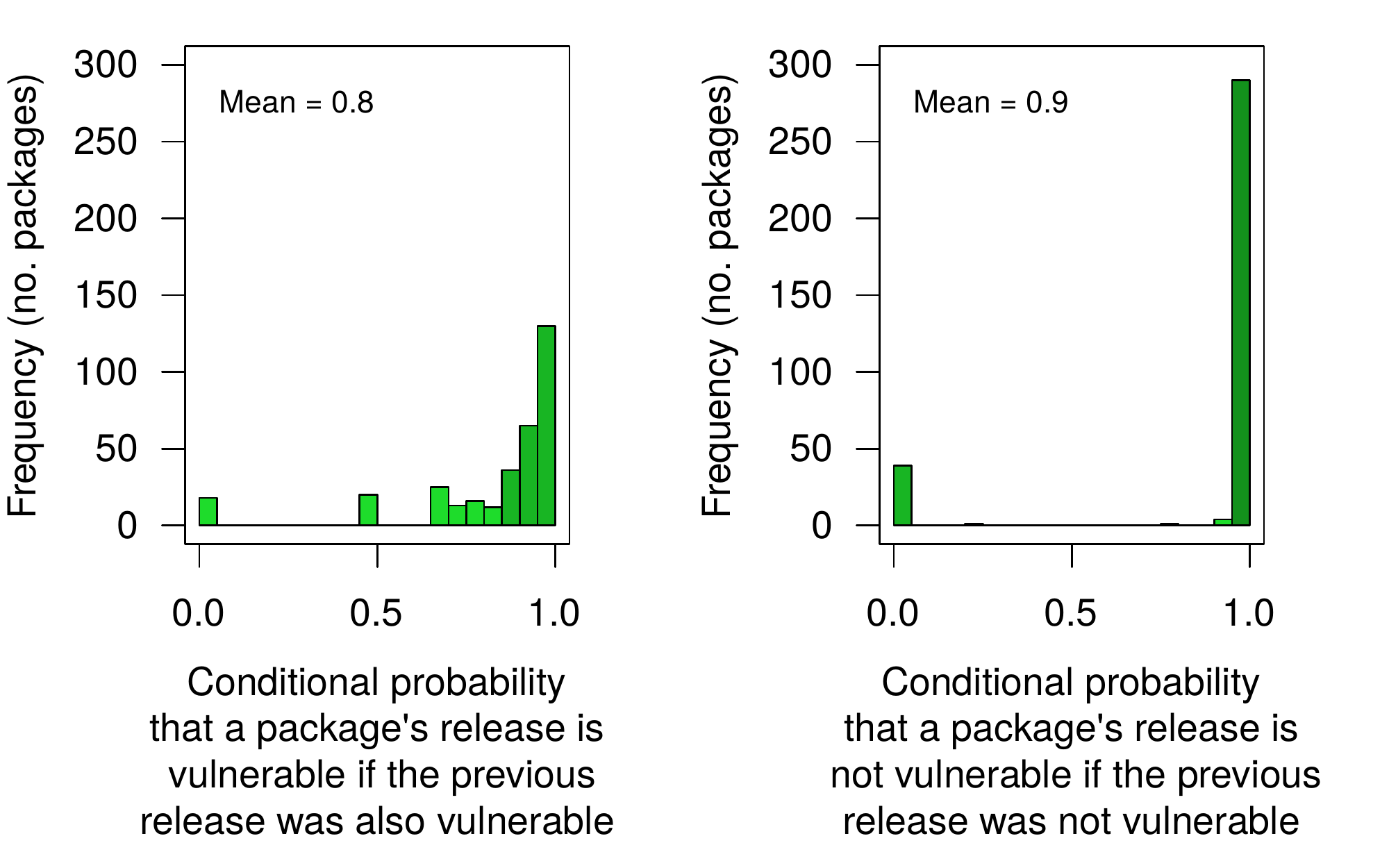}
\caption{Two Conditional Probabilities (first-order Markov chains)}
\label{fig: cond}
\end{figure}

However, the unconditional probability is not ideal because there is additional information available even with the very limited data provided in Safety~DB. This information relates to sequences of releases: if a release is vulnerable, it may be probable that either the past or the future release is also vulnerable, or both are. A classical first-order Markov chain can be used to examine such conditional probabilities. Because there are only two states (vulnerable and not vulnerable), the computation amounts to iterating over all possible release pairs and then forming a two-by-two contingency table from which the transition probabilities can be calculated. The results are shown in Fig.~\ref{fig: cond} for two of the main state change trajectories.

The left-hand side plot summarizes the transition probabilities from vulnerable releases to further vulnerable releases. The result is clear: if a random vulnerable release is picked from a given package, it is highly probable that also the subsequent release is affected either by the same vulnerability or a different vulnerability. The result stems partially from the fact that most Python vulnerabilities have affected all previous releases. In other words, the condition \texttt{"<1.4.18"} in the snippet shown in Subsection~\ref{subsec: sources} is fairly typical. Because all release pairs are counted, such conditions alone are not a sufficient explanation; there is also a fair amount of transitions from vulnerable to non-vulnerable releases. Indeed, the right-hand side plot in Fig.~\ref{fig: cond} seems to indicate that once a non-vulnerable release is reached, it is very probable that this release is followed by another vulnerability-free release. A~further point is that the first-order Markov chain may not be enough to capture the longer software evolution history across releases~\cite{Ruohonen17APSEC}. To examine this problem a little further, a more formal but still simple time series analysis can be considered.

\subsubsection{Forecasts}

The binary-valued vectors can be also modeled with univariate ``autologistic'' regression models, which have been studied from the 1950s onward \cite{Cox58, Rydberg03, SaikkonenKauppi08}. In the context of time series analysis the essence behind the models is simple: the standard logistic regression formulation is augmented with a predefined number of autoregression (AR) terms. Thus, the estimated conditional probability $\hat{p}_{i_k}$ for the $i$:th release to be vulnerable in the $k$:th package is given by
\begin{equation}\label{eq: autologistic}
\Pr(w_{i_k} = 1~\vert~\textmd{past})
= \frac{
\exp(\beta_0 + \sum^{\ell_k}_{j=1}\beta_j w_{{(i-j)}_k})
}{
1 + 
\exp(\beta_0 + \sum^{\ell_k}_{j=1}\beta_j w_{{(i-j)}_k})
} ,
\end{equation}
where $\beta_0$ is a constant and $\beta_1, \ldots, \beta_{\ell_k}$ are regression coefficients. Note that the length of an estimated vector $\vcw_k$ is also reduced by the package-specific $\textmd{AR}(\ell_k)$-order of the process.

The in-sample forecast experiment is implemented by predicting the vulnerability probabilities for $t = 5$ and $t = 10$ releases, using the last five and ten releases as the forecasting targets. Two restrictions are imposed for statistical reasons: only packages with at least $r_k \geq 25$ releases are qualified to the experiment, and, furthermore, those packages are excluded for which the standard deviation of the \text{$(r_k - [t + \ell_k])$-length} training data is less than $0.25$. These restrictions ensure that all optimizations and statistical routines compute without problems. In total, $99$ packages passed the two criteria. 

To determine the AR-orders, the maximum orders were first restricted to $\alpha_k = \lfloor 0.1 \times r_k \rfloor$, and then separate autologistic models were estimated for each of the ninety-nine packages using the full samples with $\ell_k = 1, \ldots, \alpha_k$. According to the minimum values of the Akaike's information criterion (AIC), the first-order AR process was preferable for as many as $96$ packages. Two packages settled for AR(2) and one package for AR(14). Therefore, the classical Markov property seems to be sensible for the clear majority of the packages observed.

\begin{table}[th!b]
\centering
\caption{A Summary of Autologistic Forecast Performance}
\label{tab: forecast}
\begin{tabular}{lcccrccc}
\toprule
\multicolumn{3}{c}{$t = 5$} 
&& \multicolumn{3}{c}{$t = 10$} \\
\cmidrule{1-3}\cmidrule{5-7}
Mean & Median & Max && Mean & Median & Max \\
\cmidrule{1-3}\cmidrule{5-7}
$0.008$ & $0.006$ & $0.018$ && $0.014$ & $0.015$ & $0.033$ \\
\bottomrule
\end{tabular}
\end{table}

Basic model validation techniques (such as conventional cross-validation) are unsuitable in the time series context. Therefore, to summarize the overall forecasting performance with the AIC-selected orders, the absolute differences between the actual (binary) values and the predicted probabilities can be used just like in conventional time series analysis. The between-package averages of the mean, median, and maximum absolute differences are shown in Table~\ref{tab: forecast}. To further clarify,
\begin{equation}
\frac{1}{99}\sum^{99}_{k = 1}
\left(
\frac{1}{t} \sum^{r_k}_{i=s}
\vert w_{i_k} - \hat{p}_{i_k} \vert
\right) ,
~s = r_k - (t + \ell_k) + 1 ,
\end{equation}
is used for the mean columns in the table, for instance. According to the numbers shown, the predictions mostly based on AR(1) models are highly sensible. The averages of the maximum absolute differences are below $0.04$ in both forecasting windows, for instance. Because most of the releases in the $t = 5$ and $t = 10$ forecast windows have not been vulnerable, the results largely reinforce the right-hand side plot in Fig.~\ref{fig: cond}, given that $\Pr(w_{i_k} = 0~\vert~\textmd{past}) = 1 - \hat{p}_{i_k}$ via \eqref{eq: autologistic}. If the conventional $\hat{p}_{i_k} \geq 0.5$ threshold is used to assign the predicted probabilities to the binary categories of vulnerable and non-vulnerable releases, the average accuracy (correct predictions to all predictions) is as high as $0.99$. To adjust the interpretation, consider a na\"ive predictor that simply assigns all values in a package's test data according to whether majority of the releases in the package's training data have been either vulnerable or non-vulnerable. The average accuracy across packages is as low as $0.42$ for such a na\"ive predictor. As is soon discussed, the exceptionally good accuracy rates for the autologistic models do not tell the whole story, however.

\section{Discussion}

This paper examined vulnerabilities in common Python packages. The results can be summarized with three points. By assuming that the CVE-referenced subset generalizes to all vulnerabilities in Safety~DB, the first point is that many of the vulnerabilities observed have been only mildly severe. Input validation and XSS have been the most typical weaknesses behind the vulnerabilities. Given that XSS is classified as a moderately or even highly severe vulnerability in related repository-specific vulnerability databases~\cite{Decan18a}, the observation bespeaks about the desirability of using standardized frameworks such as CVSS and CWE for practical vulnerability tracking. The result also underlines that Safety~DB is implicitly biased toward Python packages used for web development. By implication, the potential generalizability of the paper's results to all Python packages should be taken with a grain of salt. 

Second, the uniform distribution seems to describe relatively well the between-package unconditional probabilities for the packages' releases to be vulnerable. If a random package from Safety~DB would be picked for deployment, the choice would be close to random in terms of the unconditional probability for the package's releases to be vulnerable compared to the releases of other packages. However, such random picking is obviously not a realistic deployment scenario. For further research with practical relevance, a good question would be the conditional probability for a whole container to be vulnerable given the packages deployed within the container. The question is challenging because the univariate per-package conditional probabilities are dependent on release histories.

Thus, third, the time series results indicate that only the recent past seems to be relevant for univariate estimation of conditional probabilities. While this result is partially explained by the fact that most vulnerabilities in Python packages have affected all previous releases of the packages, the result is also sensible from a practical package management viewpoint---it is the currently installed release that sets the reference point for an upgrade. Besides supporting somewhat similar observations made earlier \cite{Murtaza16}, the result aligns with the more general software evolution research domain within which only short autoregressive orders are often observed to perform well~\cite{Rasool17}. All in all, the classical Markov property seems sensible according to the results. By implication, practical foresight based on release histories may be difficult despite of the good statistical performance reported. If only a current release is used to predict whether a future release is vulnerable, the prediction is arguably haphazard in terms of actual security: the probability may be low, but vulnerabilities still do occur.

Moreover, another question is whether there are cross-sectional (between-package) correlations \cite{Ruohonen17APSEC}. In other words, the probability that a package's release is vulnerable may be conditional on whether a release of some other package is either vulnerable or has been vulnerable in the past. This reasoning points toward the analysis of dependencies between packages. Although good progress has recently been made with continuous (calendar) time approaches \text{\cite{Decan18b, Paschenko18}}, potential dependencies between discrete-time state changes remain poorly understood. To further complicate things, state changes should be examined on the side of deployments; a package is merely software---the potential security risks realize only when the package is actually installed somewhere. For instance, a recent industry case study indicated that many of the vulnerable dependencies were not actually deployed; hence, the security risks were limited \cite{Paschenko18}. Further research is required to examine similar questions on a larger scale. Indeed, Safety~DB could be used in Internet measurement research for examining vulnerable web applications deployed in the wild.

Finally, a few words are warranted about Safety~DB and related databases in the context of practical vulnerability tracking. The recent challenges in the global CVE-based tracking infrastructure \cite{LiPaxson17, Ruohonen18IST} have also increased the proliferation of vulnerability databases. These challenges motivated also the introduction of Safety~DB~\cite{Pyup18}. Yet, in the big picture a centralized tracking infrastructure is desirable for software vulnerabilities. To this end, further research is required also regarding the means by which the challenges could be resolved. This line of research involves the practical question of how the repositories and the repository-specific databases could be integrated into the centralized CVE-based infrastructure. The topic is timely; there have been talks~\cite{MITRE18d} about the possibility to assign CVEs via GitHub for projects hosted on the platform.

\bibliographystyle{IEEEtran}


\end{document}